\documentclass[twocolumn,RNAAS]{aastex63}

\usepackage{color}
\usepackage{enumitem}
\usepackage{hyperref}
\usepackage{verbatim}
\usepackage{amsmath,amstext,mathrsfs}
\usepackage[all]{hypcap} 
\usepackage{afterpage}    
\usepackage{marginnote}
\usepackage{makecell}
\usepackage{subfigure}

\shorttitle{Modeling of Galactic Foreground Polarization with Velocity Gradients}
\shortauthors{Hu \& Lazarian}
\graphicspath{{./}{figures/}}

\begin{document}

\title{Modeling of Galactic Foreground Polarization with Velocity Gradients}

\email{yue.hu@wisc.edu, alazarian@facstaff.wisc.edu}

\author[0000-0002-8455-0805]{Yue Hu}
\affiliation{Department of Physics, University of Wisconsin-Madison, Madison, WI 53706, USA}
\affiliation{Department of Astronomy, University of Wisconsin-Madison, Madison, WI 53706, USA}
\author{A. Lazarian}
\affiliation{Department of Astronomy, University of Wisconsin-Madison, Madison, WI 53706, USA}



\begin{abstract}
The detection of primordial B-mode polarization is still challenging due to the relatively low amplitude compared to the galactic foregrounds. To remove the contribution from the foreground, a comprehensive picture of the galactic magnetic field is indispensable. The Velocity Gradient Technique (VGT) is promising in tracing magnetic fields based on the modern understanding of the magneto-hydrodynamic turbulence. In this work, we apply VGT to an \ion{H}{1} region containing an intermediate velocity cloud and a local velocity cloud, which are distinguishable in position-position-velocity space. We show that VGT gives an excellent agreement with the Planck polarization and stellar polarization. We confirm the advantages of VGT in constructing the 3D galactic magnetic field. 

\end{abstract}
\keywords{Interstellar medium (847); Interstellar magnetic fields (845); Interstellar dynamics (839)}


\section{The Velocity Gradients Technique}
\label{sec:intro}

As an output of modern MHD turbulence theories, the Velocity Gradients Technique (VGT) has already been developed as a new method to trace the magnetic fields \citep{LY18a}. VGT drives the revolutionary understanding of the galactic magnetic fields. For instance, \citet{EB} and \citet{2020MNRAS.tmp.1740L} utilize VGT and the \ion{H}{1} emission line to predict the foreground dust polarization. The ratio of the E- and B-modes calculated from VGT is BB/EE $\approx$ 0.53 $\pm$0.10, which agrees with the result from Planck polarization measurement \citep{EB}. 

The VGT approach was used in  \cite{2019ApJ...874...25G} to obtain the first 3D map of the galactic disk plane-of-the-sky magnetic field distribution. The authors successfully tested this  distribution by predicting the polarization from a number of stars with known distances and comparing them with the measured polarization. 
A similar approach, the Rolling Hough Transform (RHT; \citealt{2014ApJ...789...82C}), was applied later in \citet{2019ApJ...887..136C} to find the mean magnetic field directions in an intermediate velocity cloud (IVC) and a local velocity cloud (LVC), which are distinct in velocity space along the same LOS \citep{2019ApJ...872...56P}. In this paper, we apply VGT to the same clouds and compare the results. This is an illustration of our claim that everything possible to be obtained with RHT is available with VGT.

\begin{figure*}[t]
    \centering
    \includegraphics[width=1.00\linewidth,height=0.69\linewidth]{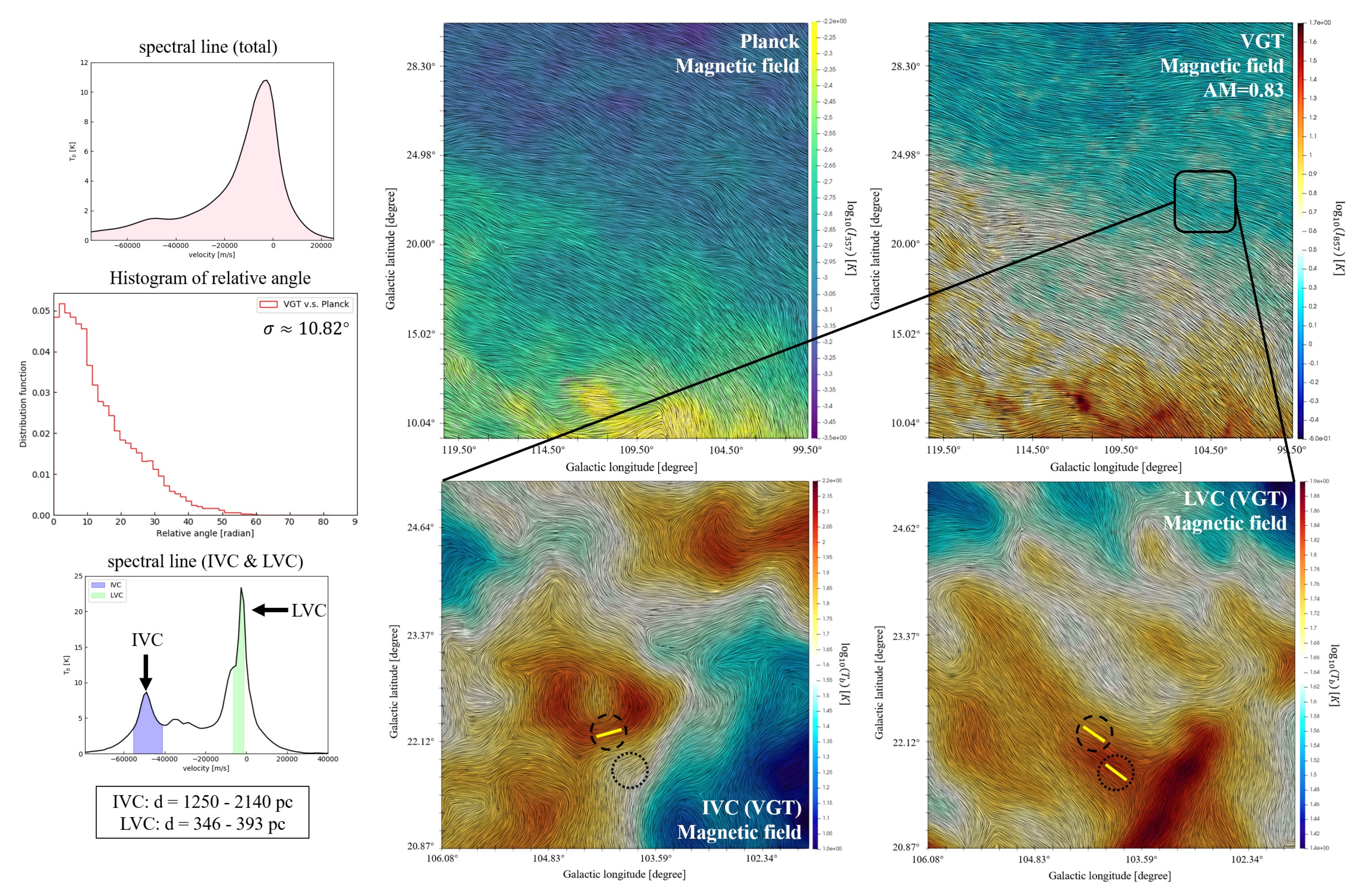}
    \caption{\label{fig:VGT} \textbf{Right}: the morphology of the magnetic fields inferred from Planck polarization (top left) and VGT (top right and bottom). \textbf{Left:} the spectral line used for the calculation and also the histogram of the relative angle between the Planck polarization and VGT. The yellow segments indicate the direction of stellar polarization. The radius of dashed circles is $\approx 0.4^\circ$ for visulization purpose. 
    }.
\end{figure*}
\citet{LP00} predicted the intensity fluctuations in PPV cubes could arise due to turbulent velocities along the LOS, which is called the velocity caustics effect.
Based on this theory, \citet{LY18a} proposed that velocity gradients in thin velocity channels can trace the POS magnetic field. Here we also employ this concept and extract the velocity gradient from all thin channels. A detailed recipe can be found in \citet{PCA,EB,Hu20}

As a separate development,  \citet{2014ApJ...789...82C} observationally found the alignment of filaments in the channel maps with the magnetic field. This empirical alignment was argued in \citet{2019ApJ...874..171C} to be related to the two-phase nature of \ion{H}{1}, while the role of velocity caustics was totally ignored. The empirical way of tracing filaments, i.e., the RHT-technique, was used to trace the magnetic field and to predict the dust polarization using \ion{H}{1} emission \citep{2015PhRvL.115x1302C,2019ApJ...887..136C}.

Compared to empirical RHT, VGT is based on the foundations routed in the anisotropic properties of MHD turbulence and the theory of turbulent reconnection. The ability of VGT in tracing magnetic fields does not depend on the media being one or two-phase, which was demonstrated numerically in \cite{IGs} and by application to molecular clouds \citep{survey,velac}. 

Due to the existence of two interpretations, the nature of fluctuations in channel maps became a subject of debates. The arguments in favor of them being pure density features are provided in \citet{2019ApJ...874..171C}. The counter-arguments stressing the role of velocity fluctuations are provided by \citet{2019arXiv190403173Y}, who appeals to two decades of theoretically and numerically studies. The velocity caustics effect must inevitably present in the thin velocity channel maps, as a natural effect of non-linear spectroscopic mapping.

In terms of the foreground polarization, \cite{EB} and \citet{2020MNRAS.tmp.1740L} demonstrated that VGT has better performance compared to RTH. In addition, VGT, as predicted in  \citet{2018ApJ...865...46L}, can provide the distribution of media magnetization of the media, as demonstrated e.g. in \citet{survey}.


\section{Results} 
\label{sec:result}
We use the \ion{H}{1} emission line from the HI4PI survey with spectral resolution $\Delta v$ = 1.49 km/s \citep{2016A&A...594A.116H}. 
We apply VGT to all thin channels of the \ion{H}{1} emission in the velocity range of -75 $<v<$ 25 km/s. Here we rotate the resulting gradients by 90$^\circ$ to indicate the magnetic field orientation $\psi$. The result is shown in Fig.~\ref{fig:VGT}. We make comparisons with the Planck 353 GHz polarized dust signal data \citep{2018arXiv180706207P}. 
The relative orientation between $\psi$ and $\phi$ is quantified by the Alignment Measure (AM). AM = 1 means a  perfect alignment case.  Here, the resulting AM is 0.83.
Also, we plot the normalized histogram of the relative angle between $\psi$ and $\phi$ (see Fig.~\ref{fig:VGT}). The histogram is close to a Gaussian distribution with the standard deviation $\sigma\approx 10.82^\circ$. Therefore, we can conclude that VGT gives an excellent agreement with Planck. 

In addition, we apply VGT to the IVC and LVC identified by \citet{2019ApJ...872...56P} centering on ($l$, $b$) = (104.08$^\circ$, 22.31$^\circ$). In particular, \citet{2019ApJ...872...56P} find the LVC locates at a distance of 346 - 393 pc associated with \ion{H}{1} emission in the velocity range of -3.8 $<v<$ -1.2 km/s. As for the IVC, it locates at a distance of 1250 - 2140 pc and -55 $<v<$ -41 km/s. The \ion{H}{1} emission within these velocity ranges are used for the calculation, respectively for IVC and LVC. The resulting magnetic field morphology is shown in Fig.~\ref{fig:VGT}. Here we make comparison with the stellar polarization centered on ($l$, $b$) = (103.90$^\circ$ , 21.97$^\circ$) and ($l$, $b$) = (104.08$^\circ$, 22.31$^\circ$) associated with these clouds. The measured mean magnetic field from stellar polarization over a 0.16$^\circ$ circle is $\langle \phi_*\rangle$ = 106$\pm$8$^\circ$ for IVC and $\langle \phi_*\rangle$ = 42.6$\pm$1$^\circ$ for IVC \citep{2019ApJ...872...56P}. We compute the mean magnetic field orientation inferred from VGT over the same region. We find $\langle \psi \rangle$ = 106.3$^\circ$ for the IVC and $\langle \psi \rangle$ = 43.5$^\circ$ for the LVC, which agree with the results of stellar polarization, as well as \citet{2019ApJ...887..136C}, where the authors got $\langle \psi_{RHT} \rangle^{IVC}$ = 111.6$^\circ$ and $\langle \psi_{RHT} \rangle^{LVC}$ = 42.6$^\circ$. However, the output of RHT depends on three parameters as inputs: a smoothing kernel diameter, window diameter, and intensity threshold  (\citealt{2014ApJ...789...82C,2015PhRvL.115x1302C}). In this sense, VGT is parameter-free and provides more statistical information, making our estimates of polarization more robust. Compared to RHT, VGT is able to trace magnetization \citep{survey} with additional advantages listed in \citet{2020MNRAS.tmp.1740L}.


\acknowledgments

{Y.H. and A.L. acknowledges the support of the NASA TCAN 144AAG1967, the NSF grant AST 1715754, and 1816234.}







\end{document}